\documentclass[conference,10]{IEEEtran}
\usepackage{calc}

\usepackage{xcolor,amsmath}
\usepackage[normalem]{ulem}
\usepackage{url}
\makeatletter
\DeclareUrlCommand\ULurl@@{%
  \def\UrlLeft{\uline\bgroup}%
  \def\UrlRight{\egroup}}
\def\ULurl@#1{\hyper@linkurl{\ULurl@@{#1}}{#1}}
\DeclareRobustCommand*\ULurl{\hyper@normalise\ULurl@}

\renewcommand{\baselinestretch}{0.9}

\usepackage{cite,amssymb,amsmath,psfrag,subfigure}
\usepackage{graphicx}
\usepackage{amssymb}
\usepackage{amsmath}
\usepackage[displaymath,mathlines]{lineno}
\usepackage{cite}
\usepackage{subfigure}
\usepackage{nicefrac}
\usepackage{color}
\usepackage{tabulary}
\usepackage{multirow}
\usepackage{mathtools}
\mathtoolsset{showonlyrefs}

\usepackage{bm}
\newcommand{\mbs}[1]{\bm{#1}}
\newcommand{\mat}[1]{{\uppercase{\mbs{#1}}}}
\newcommand{\Id}{\mat{\mathrm{I}}}

\def\nn{\nonumber}
\renewcommand{\H}{{\scriptscriptstyle\mathsf{H}}}

\DeclareMathAlphabet{\mathpzc}{OT1}{pzc}{m}{it}

\newtheorem{proposition}{Proposition}

\newtheorem{remark}{Remark}

\DeclareMathOperator{\tr}{\mathrm{tr}}

\DeclareMathOperator{\M}

\newcommand{\CG}[2]{\mathcal{CN}\left({#1},{#2}\right)}

\newcommand{\B}[1]{{\pmb{#1}}}

\newcommand{\T}{\mathcal{T}}

\newcommand{\bo}[1]{{\pmb{#1}}}
\newcommand{\al}{\alpha}

\IEEEoverridecommandlockouts
\begin{document}

\title{Impact of User Mobility on Optimal Linear Receivers in Cellular Networks}
\author{Anastasios K. Papazafeiropoulos \\
Communications and Signal Processing Group, Imperial College London, London, U.K.\\
Email: a.papazafeiropoulos@imperial.ac.uk
\thanks{This research was supported by a Marie Curie Intra European Fellowship within the $7$th European Community Framework Programme for Research of the European Commission under grant agreement no. [$330806$], IAWICOM.}}

\maketitle

%%%%%%%%%%%%%%%%%%%%%%%%%%%%%%%%%%%%%%%%%%%%%%%%%%%%%%%%%%%%%%%%%%%%%
\vspace{-1cm}
%
% \begin{abstract}
%
% \end{abstract}
%
% \begin{keywords}
% \end{keywords}

\begin{abstract}
 We consider the uplink of non-cooperative multi-cellular systems deploying multiple antenna elements at the base stations (BS), covering both the cases of conventional and very large number of antennas. Given the inevitable pilot contamination and an arbitrary path-loss for each link, we address the impact of time variation of the channel due to the relative movement between users and BS antennas, which limits system's performance even if the number antennas is increased, as shown. In particular, we propose an optimal linear receiver (OLR) maximizing the received signal-to-interference-plus-noise (SINR). Closed-form lower and upper bounds are derived as well as the deterministic equivalent of the OLR is obtained. Numerical results reveal the outperformance of the proposed OLR against known linear receivers, mostly in environments with high interference and certain user mobility, as well as  that massive MIMO is preferable even in time-varying channel conditions.
\end{abstract}
 \section{Introduction}
 Multiple-input multiple-output (MIMO) systems have appeared as a method to increase the data rate and the spectral efficiency over wireless links through multiplexing gain and spatial diversity~\cite{Telatar}. However, the evolution of cellular networks has resulted to a greedy need for higher data rates, achieved by means of multi-user MIMO (MU-MIMO) capable of exploiting multi-user diversity~\cite{gesbert}. Despite that the channel state information at the transmitter (CSIT) contributes little to the multiplexing gain in single-user scenarios, its knowledge in MU-MIMO is crucial~\cite{Jindal}.
 
 Nevertheless, a new breakthrough technique, called massive MIMO or very large MIMO, makes it possible to apply low-complexity linear decoders to maximize system capacity by vanishing thermal noise, intra-cell interference, and estimation errors~\cite{Marzetta}. In massive MIMO cellular networks, a base station (BS) deploys a large number of antennas, and multiple users can be served by the same time-frequency resource, since more degrees of freedom are provided.
 
 Unfortunately, the reuse of pilot sequences in adjacent cells causes pilot contamination~\cite{Marzetta,Jose}, which degrades the quality of CSIT. In addition, the commonly present mobility of the users, implying time-varying channel conditions, induces extra inaccuracy at the CSIT. As a consequence,  the current CSIT cannot be perfectly known but only estimated.
 
 It has been shown that maximum-ratio combining (MRC), zero-forcing (ZF), and minimum mean-square error (MMSE) perform the same in single-cell systems, when the throughput per user is $1$ bit per channel use, but MMSE is in general optimal~\cite{Ngo_Energy}. The behavior of MMSE has been studied extensively in multi-cell networks, and in fact, in~\cite{Performance_Ngo} an optimal linear receiver (OLR), achieving the maximum signal-to-interference-plus-noise ratio (SINR), has been obtained. Nevertheless, some limited prior work on massive MIMO has considered the effect of channel aging due to user mobility. Specifically, in~\cite{Truong}, Truong et. al derived deterministic equivalents for the MRC (uplink) and maximum-ratio-transmission (MRT) for the downlink case, while in~\cite{Papazafeiropoulos1,Papazafeiropoulos2}, the more complex deterministic equivalents for MMSE (uplink) and regularized ZF (downlink) were obtained and a comparison among the various strategies was provided. In the  context of channel aging, the exact and approximate sum-rates have been obtained  in~\cite{Papazafeiropoulos1} for finite and large number of antennas, respectively, when zero-forcing (ZF) decoder is applied. The arising need is to derive an OLR that addresses the inevitable channel aging occurring in time-varying channels. 
 
 Motivated by the above discussion, we hereafter derive the OLR in a cellular scenario subject to time-varying channel conditions, exploiting the correlation between the channel estimates and the interference from other cells. Furthermore, we elaborate on a comparison among various known receivers (MRC, ZF, and MMSE) to depict the outperformance of the proposed OLR and shed light on how the user mobility affects the performance in such case. In fact, we obtain lower and upper bounds of the sum-rate, accounting for channel aging, with an OLR assuming a finite  number of BS antennas. In case of large number of antennas, we derive the deterministic equivalent approximation of the SINR that gives us the ability of avoiding lengthy Monte-Carlo simulations. Notably, it is of high importance to massive MIMO that in time-varying channels they are still favorable, but  the performance of the system is limited, even if the number of antennas is cranked up, as shown by our results. 
 
  \section{System Model}
  We consider a cellular uplink scenario, where the system is composed of $L$ cells with one multi-antenna BS ($N$ antennas) and $K$ single-antenna users $\left( N > K \right)$ per cell. All users share the same time-frequency resource, while we assume  non-cooperative BSs. Taking into account a quasi-static block fading model, where the channels are constant during the symbol period but vary from symbol to symbol, the channel vector $\B g_{lik}[n]\in\mathbb{C}^{N\times 1}$ between the $k$th user in the $i$th cell and the BS in cell $l$ at the $n$th time-slot is decomposed as independent small-scale fading by means of $\B h_{lik}\in\mathbb{C}^{N\times 1}\sim \CG{\B{0}}{\Id_N}$ and  large-scale fading (shadow  fading and path loss) by means of $\beta_{lik}$, given that the antennas are sufficiently well separated in realistic systems. More concretely, we have
\begin{align}
\B g_{lik}[n]= \sqrt{\beta_{lik}} \B h_{lik}[n].\label{eq MU-MIMO 1}
\end{align}

In simple mathematical terms, the $N\times1$ frequency-flat received signal  vector at the $l$th BS $\left( l=1, 2, ...L \right)$ at the $n$th time-slot is as follows
 \begin{align}
    \B{y}_l[n]
    &=        \sqrt{p}         \sum_{i=1}^{L} \B{G}_{li}[n]  \B{x}_i[n]   +  \B{z}_l[n]\nn\\
    &= \sqrt{p}         \sum_{i=1}^{L} \B{H}_{li}[n]  \B{D}_{li}^{{1}/{2}} \B{x}_i[n]
       ,\label{eq MU-MIMO 2}
\end{align}
where  $\sqrt{p}\, \B{x}_i[n]\in \mathbb{C}^{K\times 1}$ is the zero-mean stochastic transmit signal vector of $K$ users
allocated in the $i$th cell with $p>0$ being the
average transmitted power per user,  $\B{n}_l$ is the additive white Gaussian noise (AWGN) vector at the receiver (BS), modeled as $\CG{\B{0}}{\Id_N}$, and $\B G_{li}[n]= \big[\,\B g_{li1}[n],\ldots, \B g_{liK}[n]\big]\in \mathbb{C}^{N\times K}$ denotes the complex combined matrix between all users in cell $i$ and BS $l$. Note that in~\eqref{eq MU-MIMO 2}, we have applied~\eqref{eq MU-MIMO 1} in matrix form, i.e., where $ \B{H}_{li}[n]\in\mathbb{C}^{N\times K}$ is the fast fading matrix and $\B D_{li}$ is a $ K \times K$ diagonal matrix with $ [\B D_{li}]_{kk}=\beta_{lik}$\footnote{Regarding shadowing, i.e., $\beta_{lik}$, it is independent of symbol's index $n$ because we assume that the coherence time is large relative to any  large-scale  delay constraint of the channel.}.

The detection of the signals transmitted from $K$ users by their associated BS necessitates 
the knowledge of channel state information (CSI), which can be obtained by using uplink pilots. Unfortunately, in realistic systems, the CSI at the BS is degraded by pilot contamination. This effect arises  due to frequency reuse across a multiplicity of cells. According to MMSE channel estimation, the channel can be written as~\cite{Truong}
\begin{align}
 \B G_{li}[n] = \hat{\B G_{li}}[n] + \tilde{\B G}_{li}[n],\label{eq:MMSEorthogonality}
\end{align}
where  
$\mathrm{vec}\!\left( \hat{\B G}_{li}^{\dagger}[n] \right)=\hat{\B H}_{li}[n]\hat{ \B{D}}_{li}^{\frac{1}{2}}\!\!\! \sim \!\!\CG{\B 0}{\Id_{N}\otimes\hat{ \B{D}}_{li}}$  and $\mathrm{vec}\!\left( \tilde{\B G}_{li}^{\dagger}[n] \right) \!\!\sim \!\!\CG{\B 0}{\Id_{N}\otimes \left({\B{D}}_{li}-\hat{\B{D}}_{li}\right)}$ are the independent channel estimate and channel estimation error matrices, respectively. Note that $\hat{\B H}_{li}[n]$ is a $N\times N$ matrix with Gaussian entries having zero mean and unit variance as well as $\hat{\B{D}}_{li}$ is a $K\times K$ diagonal matrix with elements 
$[\hat{\B{D}}_{li}]_{kk}=\hat{\beta}_{lik}=\frac{\beta^{2}_{lik}}{\sum_{j=1}^{L}\beta_{ljk}+1/p_{\mathrm{p}}}$, where $p_{p}$ is the transmit power of each pilot symbol $\left( p_{p}=\tau p \right)$. Nevertheless, an important relationship, going to help our derivations, concerns the estimated channel between the $l$th BS and the users in cell $i$. It is given by~\cite{Performance_Ngo}
\begin{align}
 \hat{\B{G}}_{li}[n]=\hat{\B{G}}_{ll}[n]\B{R}_{li},\label{estimatedChannel}
\end{align}
where $\hat{\B
G}_{li}[n]\triangleq\left[\hat{\B g}_{li1}[n],\ldots, \hat{\B
g}_{liK}[n]\right]\in \mathbb{C}^{N\times K}$ is the  estimated combined
channel matrix from all users in cell $i$ to BS $l$ and $\B {{{R}}}_{li}=\mathrm{diag}\big\{\frac{{\beta}_{li1}}{{\beta}_{ll1}},\frac{{\beta}_{li2}}{{\beta}_{ll2}},\ldots,\frac{{\beta}_{liK}}{{\beta}_{llK}}
\big\}$.

Besides pilot contamination, in any common propagation scenario, a relative movement takes place between the antennas and the scatterers that degrades more channel's  performance. Under these circumstances, the channel is time-varying and can be modeled by the famous Gauss-Markov block fading model~\cite{Caire}, which is basically an autoregressive model of certain order that incorporates two-dimensional isotropic scattering (Jakes model). More specifically, our analysis targets to relate the current channel state with its
past samples. For reasons of simplicity and computational complexity, we consider the following autoregressive model of order $1$~\cite{Truong}
\begin{align}
 \B G_{li}[n]=\alpha {\B G}_{li}[n-1]+\B E_{li}[n],
\label{eq:aut}
\end{align}
where  $\mathrm{vec}\!\left( \B E_{li}^{\dagger}[n] \right)\!\sim\!
\CG{\B 0}{ \Id_{N}\otimes\left(1\!-\!\alpha^{2}\right){\B{D}}_{li}}$ and ${\B G}_{li}[n\!-\!1]$ 
are uncorrelated. They model the stationary Gaussian channel error vector due to
the time variation of the channel and the channel at the previous symbol
duration. The $\alpha\!=\!\mathrm{J}_{0}\left( 2 \pi f_{D} T_{s} \right)$ parameter, where $\mathrm{J}_{0}(\cdot)$ is the zeroth-order Bessel function of the
first kind, $f_{D}$ and $T_{s}$ are the maximum Doppler shift and the channel sampling period, expresses the two-dimensional isotropic scattering. It is worthwhile to mention that the maximum
Doppler shift $f_{D}$ equals $f_{D}=\frac{v f_{c}}{c}$, where $v$ (in m/s) is the relative velocity of the user, $c=3\times10^{8}$m/s is the speed of light, and $f_{c}$ is the carrier frequency.

The combination of~\eqref{eq:MMSEorthogonality} in~\eqref{eq:aut} allows the characterization of both effects at the same time instance, i.e., it provides the channel at time-slot $n$, accounting for pilot contamination and time variation of the channel. Specifically, we have
\begin{align}
 \B G_{li}[n]&=\alpha {\B G}_{li}[n-1]+\B E_{li}[n]\nonumber\\
&=\alpha \hat{\B G}_{li}[n-1]+ \tilde{\B E}_{li}[n],\label{eq:MMSEchannelEstimate}
\end{align}
where $\hat{\B G}_{li}[n-1]$ and $\tilde{\B {E}}_{li}[n]= \alpha \tilde{\B G}_{li}[n-1]+\B E_{li}[n]\sim \CG{\B{0}}{\Id_{N}\otimes\left({\B{D}}_{li}-\alpha^{2}\hat{\B{D}}_{li}\right)}$
are mutually independent. 

We focus on the study of linear detection by the receiver (BS) by means of a linear $N \times K$ matrix $\bo W_{l}[n]$ dependent on the channel estimate $\hat{\B{G}}_{li}[n]$ After applying the detector $\bo{W}_{l}[n]$ to its received signal $\bo y_{l}[n]$ at time-slot $n$, the $l$th BS obtains
  \begin{align}
 & \bo{r}_{l}=\bo{W}_{l}^{\H} [n]\bo{y}_{l}=\sqrt{p}\,\bo{W}_{l}^{\H}[n]\sum_{i=1}^{L}\bo{G}_{li}[n]\bo{x}_{i}[n]+\bo{W}_{l}^{\H} [n]\bo{z}_{l}[n]\nn\\
  \label{received_signal without Imp CSI}
  &=\al \sqrt{p}\,\bo{W}_{l}^{\H}[n]\sum_{i=1}^{L}\hat{\bo{G}}_{li}[n-1]\bo{x}_{i}[n]\nn\\ &+\sqrt{p}\,\bo{W}_{l}^{\H}[n]\sum_{i=1}^{L}\tilde{\bo{E}}_{li}[n]\bo{x}_{i}[n]+\bo{W}_{l}^{\H} [n]\bo{z}_{l}[n]
   \end{align}
where in~\eqref{received_signal without Imp CSI}, we have used~\eqref{eq:MMSEchannelEstimate}.
 Hence, the $k$th element of $\bo r_{l}$, corresponding to the signal from the $k$th user, is given by
\begin{align}
&r_{lk}= \al \sqrt{p}\,\bo{w}_{lk}^{\H}[n]\hat{\bo{g}}_{llk}[n-1]{x}_{lk}[n]\\ &\!+\!\al \sqrt{p}\!\sum_{j\ne k}^{K}\!\!\bo{w}_{lk}^{\H}[n]\hat{\bo{g}}_{llj}[n\!-\!1]{x}_{lj}[n]\!+\!\al \sqrt{p}\!\sum_{i\ne l}^{L}\!\bo{w}_{lk}^{\H}[n]\hat{\bo{G}}_{li}[n\!-\!1]\bo{x}_{i}[n]\nn\\
&\!+\!\sqrt{p}\sum_{i=1}^{L}\bo{w}_{lk}^{\H}[n]\tilde{\bo{E}}_{li}[n]\bo{x}_{i}[n]+\bo{w}_{lk}^{\H}\bo{z}_{l}[n],
\end{align}
where $x_{lk}$ and $\bo w_{lk}$ are the $k$th element and column of $\bo x_{l}$ and $\bo W_{l}$, respectively. Taking into account the independence between $\hat{\bo{G}}_{li}[n-1]$ and $\tilde{\bo{E}}_{li}[n]$, the achievable uplink SINR $\mathrm{SINR}_{k}$ of the $k$th user can be written as in~\eqref{general sum_rateOLR}\footnote{We assume encoding of the message over many realizations of all sources of randomness in the model including noise and channel estimate error~\cite{Performance_Ngo}.}. 
\begin{figure*}[!t]
\begin{align}
\mathrm{SINR}_{k}=\frac{\al^{2} {p}|\bo{w}_{lk}^{\H}[n]\hat{\bo{g}}_{llk}[n-1]|^{2}}{\al^{2} {p}\sum_{j\ne k}^{K}|\bo{w}_{lk}^{\H}[n]\hat{\bo{g}}_{llj}[n-1]|^{2}+{p}\sum_{i=1}^{L}\bo{w}_{lk}^{\H}[n]{\bo{R}}_{\tilde{E}}\bo{w}_{lk}[n]+\al^{2} {p}\sum_{i\ne l}^{L}\|\bo{w}_{lk}^{\H}[n]\hat{\bo{G}}_{li}[n-1]\|^{2}+\|\bo{w}_{lk}^{\H}[n]\|^{2}}.\label{general sum_rateOLR}
\end{align}
\hrulefill
\end{figure*}
Note that  $\displaystyle {\bo{R}}_{\tilde{E}}=\sum_{i=1}^{L}\mathbb{E}\left\{\tilde{\bo E}_{li} \tilde{\bo E}_{li}^{\H}\right\}$, i.e.,
\begin{align}
 {\bo{R}}_{\tilde{E}}=\sum_{i=1}^{L}\tilde{e}_{li}\Id_{N}=\sum_{i=1}^{L}\sum_{k=1}^{K}\left[ \beta_{lik}-\al^{2} \hat{\beta}_{lik} \right]\!\Id_{N}\!.
 \end{align}

\section{Optimal Linear Receiver}
In this section, following a similar procedure as in~\cite{Performance_Ngo}, we derive the OLR maximizing the received SINR. Specifically, we have
\begin{align}
&\mathrm{SINR}_{k}=\frac{|\bo{w}_{lk}^{\H}[n]\hat{\bo{g}}_{llk}[n\!-\!1]|^{2}}{\bo{w}_{lk}^{\H}[n] \bo \Xi_{k}[n\!-\!1] \bo{w}_{lk}[n] }\\\label{sum_rateOLR2}
    &\le \frac{\left\|\bo{w}_{lk}^{\H}[n]\bo \Xi_{k}^{1/2} [n\!-\!1]\right \|^{2}\left\|\bo \Xi_{k}^{-1/2} [n\!-\!1]\hat{\bo{g}}_{lllk}[n\!-\!1] \right \|^{2}}{\bo{w}_{lk}^{\H}[n] \bo \Xi_{k}[n\!-\!1] \bo{w}_{lk}[n] }\\ 
  &=\hat{\bo{g}}_{llk}^{\H}[n-1] \bo \Xi_{k}^{-1} [n\!-\!1]\hat{\bo{g}}_{llk}[n-1] ,\label{sum_rateOLR3}
\end{align}
where  we have denoted that $ \bo \Xi_{k}[n-1]=\sum_{j\ne k}^{K}\hat{\bo{g}}_{llj}[n-1] \hat{\bo{g}}_{llj}^{\H}[n-1]+\sum_{i\ne l}^{L}\hat{\bo{G}}_{li}[n-1]\hat{\bo{G}}_{li}^{\H}[n-1]+\frac{1}{\alpha^{2}}\left( \sum_{i=1}^{L}\tilde{e}_{li}+\frac{1}{p} \right)\Id_{N}$ as well as in~\eqref{sum_rateOLR2} we have used Cauchy-Schartz's inequality. According to its property, the equality holds when $\bo{w}_{lk}[n]=c\,\bo \Xi_{k}^{-1} [n-1]\bo{g}_{llk}[n-1]$ for any nonzero  $c \in \mathbb{C}$. It is important to mention that $\bo \Xi_{k} [n-1]$ can be also written as
\begin{align}
 \bo \Xi_{k} [n\!-\!1]\!=\!\sum_{i= 1}^{L}\!\hat{\bo{G}}_{li[k]}[n\!-\!1]\hat{\bo{G}}_{li[k]}^{\H}[n\!-\!1]\!+\!\sigma^2\Id_{N},\label{OLR}
\end{align}
where $\hat{\bo{G}}_{li[k]}[n-1]$ is the matrix $\hat{\bo{G}}_{li}[n-1]$ with its $k$th column removed and $\sigma^2=\frac{1}{\alpha^{2}}\!\left( \!\sum_{i=1}^{L}\!\tilde{e}_{li}\!+\!\frac{1}{p}\! \right)$.
\begin{remark}
 Given that the typical MMSE receiver is 
 \begin{align}
  \bo w_{lk}[n]\!\!=\!\!\alpha\!\bigg(\! \!\hat{\bo G }_{ll}[n\!-\!1]\hat{\bo G }_{ll}^{\H}[n\!-\!1]\!+\frac{1}{\alpha^{2}}\!\left( \bo Z_{l}\!+\!\frac{1}{p} \right)\Id_N\!\!\bigg)^{-1}\!\!\hat{\bo g}_{llk}[n\!-\!1],\nn
 \end{align}
 where the matrix $\bo Z_{l}$ is deterministic and equals to
 \begin{align}
  \bo Z_{l}&=\mathbb{E}\!\left[ \!\tilde{\bo E}_{ll}\tilde{\bo E}_{ll}^{\H}\!\right]\!\!+\!\sum_{i\ne l}^L\!\mathbb{E}\!\left[\bo {G}_{li}\bo {G}_{li}^{\H}\right]\nn\\
  &=\sum_{k}^{K}\left( \tilde{e}_{li}+\sum_{i\ne l}^{L} \beta_{lik} \right)\Id_{N},
 \end{align}
 and the expression of the typical MRC is $
  \bo w_{lk}[n]\!\!=\!\!\alpha\hat{\bo g}_{llk}[n\!-\!1]$, the SINR for the OLR and MMSE are expected to behave similarly in low interference conditions, while the proposed OLR achieves higher sum-rate in interference-limited scenarios, as expected~\cite{Performance_Ngo}. However, the main contribution rests on the exhibition of the effect of the user mobility on the OLR, which is clearly described in~\eqref{OLR}. As can be seen, the SINRs corresponding to all receivers under investigation coincide after a certain value of $\alpha$  that makes the term multiplied by $1/\alpha^{2}$ negligible.
 \end{remark}

Application of the eigenvalues decomposition to the following $N\times N$ matrix 
\begin{align}
\bo S[n-1] &=\sum_{i=1}^{L}\hat{\bo{G}}_{li[k]}[n-1]\hat{\bo{G}}_{li[k]}^{\H}[n-1]\\
&\overset {\eqref{estimatedChannel}}{=}\hat{\bo{G}}_{ll[k]}[n-1]\bo R_{l[k]}\hat{\bo{G}}_{ll[k]}^{\H}[n-1]\\
&\overset {\eqref{eq MU-MIMO 1}}{=}\hat{\bo{H}}_{ll[k]}[n-1]\hat{\bo D}_{l[k]}\hat{\bo{H}}_{ll[k]}^{\H}[n-1]\label{eigen1}
\end{align}
  with $\bo R_{l[k]}=\sum_{i=1}^{L}\bo R_{li[k]}^{2}$ and $\hat{\bo D}_{l[k]}=\sum_{i=1}^{L}\hat{\bo D}_{li[k]}^{2}$ provides the eigenvalues in terms of a diagonal matrix $\bo B[n-1]$ and the corresponding eigenvectors by means of the columns of the unitary matrix $\bo U$ as $\bo S[n-1]=\bo U^{\H}[n-1]\bo B[n-1]\bo U[n-1]$.
%   Note that $\hat{\bo{H}}_{ll[k]}[n-1]\sim \CG{\B 0}{\Id_{N}\otimes\Id_{K-1}}$.

Given that the rank of $\bo S[n-1]$ is $K-1$, the matrix $\bo B$ has the form
\begin{align}
 \bo B[n-1]=\mathrm{diag}\big[\lambda_{1}, \ldots,\lambda_{K-1},\overbrace{0,\ldots,0}^{N-K+1} \big]_{[n-1]},\label{eigenv_decomp}
\end{align}
where the first $K-1$ terms represent the respective eigenvalues of $\bo S[n-1]$ and the subscript $[n-1]$ in front of a matrix denotes its instance at the $\left(n-1  \right)$th slot. Taking advantage of the eigenvalues decomposition and~\eqref{OLR}, we can express~\eqref{sum_rateOLR3} as a sum of two independent terms as follows
\begin{align}
 &\mathrm{SINR}_{k}\!=\!\!\Big(\! \bo U[n\!-\!1]  \hat{\bo{g}}_{llk}[n\!-\!1] \!\Big)^{\H}\! \bigg[ \bo B[n\!-\!1] \! +\!\frac{1}{\alpha^{2}}\!\!\left( \sum_{i=1}^{L}\!\tilde{e}_{li}\!+\!\frac{1}{p}\!\! \right)\!\!\Id_{N} \!\bigg]^{-1} \nn\\ 
 &\times\bo U[n-1]  \hat{\bo{g}}_{llk}[n-1] \nn\\
 &=\big[ \bar{g}_{llk_{1}}^{*},\ldots, \bar{g}_{llk_{N}}^{*}\big]_{[n-1]}\nn\\
 &\times \bigg[\frac{\bar{g}_{llk_{1}}}{\lambda_{1}+\sigma^{2}},\ldots,\frac{\bar{g}_{llk_{K-1}}}{\lambda_{K-1}+\sigma^{2}},\frac{\bar{g}_{llk_{K}}}{\sigma^{2}},\ldots,\frac{\bar{g}_{llk_{N}}}{\sigma^{2}}\bigg]^{\T}_{[n-1]}\nn\\
 &=\underbrace{\sum_{j=1}^{K-1}\frac{|\bar{g}_{llk_{j}}|^{2}}{\lambda_{j}+\sigma^{2}}}_{\mathcal{I}_{1}}+\underbrace{\frac{1}{\sigma^{2}}\sum_{j=1}^{N-K+1}|\bar{g}_{llk_{j+K-1}}|^{2}}_{\mathcal{I}_{2}},\label{SINR_pdf}
 \end{align}
 where $\bo U[n-1]  \hat{\bo{g}}_{llk}[n-1]=\big[ \bar{g}_{llk_{1}},\ldots, \bar{g}_{llk_{N}}\big]^{\T}_{[n-1]}$. The multiplication of $ \hat{\bo{g}}_{llk}[n-1]$ with the unitary matrix $\bo U[n-1]$ preserves the statistical properties of the former one. Moreover, $\lambda_{j}$ for $j\in[1,K-1]$ and $|\bar{g}_{llk_{j}}|$ are independent, since $\lambda_{j}$ represents the eigenvalues of $\bo S[n-1]$, i.e.,  $\hat{\bo{G}}_{li}[n-1]\hat{\bo{G}}_{li}^{\H}[n-1]$ with the $k$th column removed. 
 
  The sum-rate $R_{k}$ of the $k$th user is written as
  \begin{align}\label{lower_bound7}
  R_{k}=\mathbb{E}\left[  \mathrm{log}_{2}\left( 1+ \mathrm{SINR}_{k}\right)\right].
  \end{align}
  Because of the arising difficulty to obtain the exact distribution of the $\mathrm{SINR}_{k}$, we resort to derive bounds of the sum-rate that prove to be very tight, as shown.
  
  \subsection{Upper Bound}
  \begin{proposition}
   The upper bound of $R_{k}$ for user $k$ in the reference cell j with OLR, accounting for user mobility, is 
   \begin{align}
 &R_{k}\!\le \!\mathrm{log}_{2} \bigg(\!\! 1\!+\!\frac{1}{ \prod_{i<j}^{K-1}\!\!\left( t_{lj}\!-\!t_{li} \right)}\!\!\!\sum_{v=1}^{K-1}\!\sum_{u=1}^{K-1}\!\!\frac{D_{lvu}}{\Gamma{\left(N\!-\!K\!+\!u\!+\!1  \right)}}\nn\\
& \times\left( \left( -1 \right)^{N-K+u-1}\left( \sigma^{2} \right)^{N-K+u}e^{\frac{\sigma^{2}}{t_{lv}}}\mathrm{Ei}\left( -\frac{\sigma^{2}}{t_{lv}} \right)\right.\nn\\
&+\!\!\!\sum_{r=1}^{N-K+u}\!\!\!\!\left( r\!-\!1 \right)!\!\left( -\sigma^{2} \right)^{N-K+u-r}\!\!\bigg)t_{lv}^{r+K-N-2}
\!+\!\frac{\left( N\!-\!K\!-\!1\right)}{\sigma^{2}} \!\bigg),\label{upperbound}
\end{align}
where $\mathrm{Ei}\left( z \right)=-\int_{-x}^{\infty}e^{-t}/t\mathrm{d}t$ is the exponential integral function~\cite[Eq.~(06.35.07.0002.01)]{Wolfram}. In addition,  $t_{lv}$ is the $v$th element of the diagonal matrix $\hat{\bo D}_{l[k]}$ and $D_{lvu}$ is the $\left( v,u \right)$th cofactor of a $\left( K-1 \right)\times \left( K-1 \right)$ matrix $\bo D_{l}$ whose $\left( i,j \right)$th entry is $\{\bo D_{l}\}_{i,j}=t_{li}^{j-1}$. 
  \end{proposition}
\IEEEproof
From~\eqref{lower_bound7}, we have
  \begin{align}
  R_{k}
  &\le   \mathrm{log}_{2}\left( 1+ \mathbb{E}\left[\mathrm{SINR}_{k}\right]\right)\nn\\
  &\overset{\eqref{SINR_pdf}}=\mathrm{log}_{2}\left( 1+ \mathbb{E}\left[\mathcal{I}_{1}+\mathcal{I}_{2}\right]\right),\label{lower_bound1}
 \end{align}
 where we have applied Jensen's inequality, since $\mathrm{log}_{2}\left(\cdot  \right)$ is a concave function.

The expectation in~\eqref{lower_bound1} can be calculated by taking the expectation of each term separately. Hence, the first term can be written as
\begin{align} \label{LB_1}
 &\mathbb{E}\left[\mathcal{I}_{1}\right]=\left( K-1\right) \mathbb{E}\left[\frac{|\bar{g}_{llk}|^{2}}{\lambda+\sigma^{2}}\right]\\ \label{LB_2}
 &=\left( K-1\right) \mathbb{E}\left[\frac{1}{\lambda+\sigma^{2}}\right]\\
 &=\frac{1}{ \prod_{i<j}^{K-1}\left( t_{lj}-t_{li} \right)}\sum_{v=1}^{K-1}\sum_{u=1}^{K-1}\frac{D_{lvu}}{\Gamma{\left(N-K+u+1  \right)}}\\
 &\times\int_{0}^{\infty}\!\!\!\frac{\lambda^{N-K+u}e^{-\frac{\lambda}{t_{lv}}}t_{lv}^{K-N-2}}{\lambda\!+\!\sigma^{2}}\mathrm{d}\lambda,\label{LB_3}\\
 &=\frac{1}{ \prod_{i<j}^{K-1}\left( t_{lj}-t_{li} \right)}\sum_{v=1}^{K-1}\sum_{u=1}^{K-1}\frac{D_{lvu}}{\Gamma{\left(N-K+u+1  \right)}}\nn\\
& \times\left( \left( -1 \right)^{N-K+u-1}\left( \sigma^{2} \right)^{N-K+u}e^{\frac{\sigma^{2}}{t_{lv}}}\mathrm{Ei}\left( -\frac{\sigma^{2}}{t_{lv}} \right)\right.\nn\\
&+\left.\sum_{r=1}^{N-K+u}\left( r-1 \right)!\left( -\sigma^{2} \right)^{N-K+u-r}\right)t_{lv}^{r+K-N-2},\label{lowerboundFirstTerm}
\end{align}
 where in~\eqref{LB_1}, we have used that the terms of the sum are i.i.d. Given the independence between  $|\bar{g}_{llk}|$ and $\lambda$ and that $|\bar{g}_{llk}|^{2}$ is an exponential variable with unit mean and unit variance, we lead to~\eqref{LB_2}. Keep in mind that the $K-1$ nonzero eigenvalues of $\bo S[n-1]$ have the same distribution as the N eigenvalues of $\bo S^{\H}[n-1]$. Specifically, substituting the PDF of the unordered eigenvalue of $\bo S^{\H}[n-1]$, obtained by~\cite{Jin} into~\eqref{LB_2},~\eqref{LB_3} is obtained. The  integral in~\eqref{LB_3} is solved by using~\cite[Eq.~(3.353.5)]{GR:07:Book}. Concerning the second term in~\eqref{lower_bound1}, we have, following the same way, that
\begin{align}
  \mathbb{E}\left[\mathcal{I}_{2}\right]&=\frac{\left( N-K-1\right)}{\sigma^{2}} \mathbb{E}\left[{|\bar{g}_{llk}|^{2}}{}\right]=\frac{\left( N-K-1\right)}{\sigma^{2}}.\label{LB_4}
\end{align}
Substitution of~\eqref{lowerboundFirstTerm} and~\eqref{LB_4} into~\eqref{lower_bound1} concludes the proof.\endIEEEproof\vskip -29mm
\subsection{Lower Bound}
 \begin{proposition}
   The lower bound of $R_{k}$ for user $k$ in the reference cell j with OLR, accounting for user mobility, is given by 
   \begin{align}
 &R_{k}\!\ge \!\mathrm{log}_{2} \bigg(\!\! 1\!+2\left( K-1 \right)\exp\Big[-2 \gamma -\frac{1}{ 2\prod_{i<j}^{K-1}\left( t_{lj}-t_{li} \right)}\\
&\!\times \! \!\sum_{v=1}^{K-1}\sum_{u=1}^{K-1}\frac{D_{lvu}t_{lv}^{K-N-2}\!\left( N\!-\!K\!+\!u \right)!}{\Gamma{\left(N-K+u+1  \right)}} \nn\\
  &\!\times\!\!\bigg(\!\! \mathrm{ln}\!\left( \!\sigma^{2}\! \right)\!\!\left(\! \frac{\sigma^{2}}{t_{lv}} \!\right)^{\!\!N-K+u+1}\!e^{\frac{\sigma^{2}}{t_{lv}}} 
  \!+\!\!\sum_{r=0}^{N-K+u}\!\mathrm{Ei}_{r+1}\!\!\left( \frac{\sigma^{2}}{t_{lv}} \right)\!\!\!\bigg)\!\Big]\!\bigg),\label{Upper_bound}
\end{align}
where $\gamma$ is Euler's constant with numerical value $\gamma \simeq 0.577216$ and $\mathrm{Ei}_{n}\left( z \right)=\int_{1}^{\infty}e^{-zt}/t^{n}\mathrm{d}t$, $n = 0, 1, 2, \ldots,$ $\mathrm{Re}(z) > 0$, is the exponential integral function of order $n$~\cite[Eq.~(06.34.02.0001.01)]{Wolfram}, which can be related to the  incomplete gamma function $\left( \mathrm{Ei}_{n}\left( z \right)=x^{n-1}\Gamma\left( 1-n,x \right) \right)$.
  \end{proposition}
\IEEEproof
The proof relies on the general bounding technique according to which we can re-express~\eqref{lower_bound7}, according to
\begin{align}
 R_{k}&=\mathbb{E}\big[  \mathrm{log}_{2}\left( 1+ \mathrm{exp}\big( \mathrm{ln}\left( \mathcal{I}_{1}+\mathcal{I}_{2} \right) \big)\right)\big]\nn\\ 
 &\ge\mathbb{E}\bigg[  \mathrm{log}_{2}\left( 1+ \mathrm{exp}\Big(\big( \mathrm{ln}\left( 2\mathcal{ I}_{1}\right)+ \mathrm{ln}\left(2 \mathcal{ I}_{2}\right)\big)\Big)/2\right)\bigg]\nn\\
 &\ge  \mathrm{log}_{2}\bigg( 1+ \mathrm{exp}\Big( \mathbb{E}\big[\big( \mathrm{ln}\left( 2\mathcal{ I}_{1}\right)+ \mathrm{ln}\left(2 \mathcal{ I}_{2}\right)\big)\big]/2\Big) \bigg)\!,\label{upper2}
\end{align}
 where first we have used the arithmetic mean-geometric mean inequality, and then we have exploited the fact that $\mathrm{log}_{2}\left( 1+\mathrm{exp}\left( x \right) \right)$ is convex in $x$ and thereafter  applying Jensen's inequality. The last expression is constituted by two parts. Regarding the first part, we have
 \begin{align}
  &\mathbb{E}\big[ \mathrm{ln}\left(2 \mathcal{I}_{1}\right)\!\ge\! \mathbb{E}\Big[\!\sum_{j=1}^{K-1}\!\mathrm{ln}\left(\!\frac{ 2 \left( K\!-\!1 \right)\!|\bar{g}_{llk_{j}}|^{2} }{\lambda_{j}+\sigma^2}\right)\!\!\Big / \! \!\left( K\!-\!1 \right)\Big] \label{upper3}\\
  &=\mathbb{E}\Big[\mathrm{ln}\left( \frac{2\left( K-1 \right) |\bar{g}_{llk}|^{2}}{\lambda+\sigma^2} \right)\nn\\
  &=\mathbb{E}\Big[\mathrm{ln}\left( {2\left( K-1 \right) |\bar{g}_{llk}|^{2}} \right) -\mathbb{E}\Big[\mathrm{ln}\left( {\lambda+\sigma^2}\right)\Big]\nn\\ \label{upper4}
  &=\int_{0}^{\infty}\mathrm{ln}\left( 2{\left( K-1 \right) x e^{-x}} \right)\mathrm{d}x
  -\frac{1}{ \prod_{i<j}^{K-1}\left( t_{lj}-t_{li} \right)}\nn\\
  &\times\sum_{v=1}^{K-1}\sum_{u=1}^{K-1}\frac{D_{lvu}t_{lv}^{K-N-2}}{\Gamma{\left(N-K+u+1  \right)}}\nn\\
  &\times \int_{0}^{\infty}\mathrm{ln}\left( {\lambda+\sigma^2}\right)\lambda^{N-K+u}e^{-\frac{\lambda}{t_{lv}}}\mathrm{d}\lambda\\
  &=\mathrm{ln}\left( 2\left( K\!-\!1 \right) \right)\!-\!2\gamma\!\nn\\
  &-\frac{1}{ \prod_{i<j}^{K-1}\left( t_{lj}-t_{li} \right)}
  \sum_{v=1}^{K-1}\sum_{u=1}^{K-1}\frac{D_{lvu}t_{lv}^{K-N-2}\!\left( N\!-\!K\!+\!u \right)!}{\Gamma{\left(N-K+u+1  \right)}} \nn\\
  &\!\times\!\bigg(\!\! \mathrm{ln}\left( \sigma^{2} \right)\!\!\left(\! \frac{\sigma^{2}}{t_{lv}} \!\right)\!^{N-K+u+1}e^{\frac{\sigma^{2}}{t_{lv}}} 
  \!+\!\sum_{r=0}^{N-K+u}\!\mathrm{Ei}_{r+1}\left( \frac{\sigma^{2}}{t_{lv}} \right)\!\!\bigg)\!.\label{upper5}
 \end{align}
Note that in~\eqref{upper3}, we have used again the arithmetic mean-geometric mean inequality and in~\eqref{upper4} the fact that $|\bar{g}_{llk}|^{2}$ is exponential with unit mean as well as that $\lambda$ is an unordered eigenvalue of semi-correlated Wishart distribution according to~\cite{Jin}. The last expression is obtained after using~\cite[Eq.~(4.332.1), (3.351.3)]{GR:07:Book} and~\cite[Eq.~(47]{Shin}.   Similarly, $\mathbb{E}\big[ \mathrm{ln}\left( \mathcal{I}_{2}\right)\big]$ is obtained as
\begin{align}
 \mathbb{E}\big[ \mathrm{ln}\left(2 \mathcal{I}_{2}\right)\big]=-2\gamma+\mathrm{ln}\left( 2\left( K-1 \right) \right),
\end{align}
which when substituted with~\eqref{upper5} into~\eqref{upper2}, returns~\eqref{Upper_bound}.\endIEEEproof
\section{Deterministic Equivalent of SINR}
The emerging technology of very large MIMO leads us to study the effect of finite Doppler shift on their performance. Specifically, the following proposition quantifies the degradation of the SINR due to the relative movement of users.
\begin{proposition}
   The deterministic equivalent of $\mathrm{SINR}_{k}$ for user $k$ in the reference cell j with OLR, accounting for user mobility, is given by 
   \begin{align}\label{detequiv}
 \overline{\mathrm{SINR}}_{k}{\asymp }\tr  \bo T_{li}  \hat{\beta}_{llk} ,
\end{align}
where 
\begin{align}
  \bo T_{li}= \left( \sum_{k=1}^{K} \frac{\hat{ \beta}_{lik}}{1+\delta_{lik}}+\frac{1}{\alpha^2}\left( \sum_{i=1}^{L}\tilde{e}_{li}+\frac{1}{p} \right) \right)^{-1}\Id_{N},
\end{align}
and $\delta_{lik}=\lim_{t \to \infty}\delta_{lik}^{\left( t \right)}$, where for $t=1,2,\ldots$
\begin{align}
 \delta_{lik}^{\left( t \right)}=\left( \sum_{j=1}^{K}\frac{\hat{\beta}_{lij}}{1+\delta^{\left( t-1 \right)}_{lij}}+\frac{1}{\alpha^2}\left( \sum_{i=1}^{L}\tilde{e}_{li}+\frac{1}{p} \right) \right)^{-1}\hat{ \beta}_{lik}N
\end{align}
with initial values $\delta_{lik}^{\left( 0 \right)}=\alpha^2\big /\left( \sum_{i=1}^{L}\tilde{e}_{li}+\frac{1}{p} \right) $ for all $k$.
  \end{proposition}
\IEEEproof
It is obtained by means of~\eqref{sum_rateOLR3} as
\begin{align}\label{deterministicequivalent1}
\overline{\mathrm{SINR}}_{k}&= \hat{\bo{g}}_{llk}^{\H}[n-1] \bo \Xi_{k}^{-1} [n] \hat{\bo{g}}_{llk}[n-1]\nn\\
&\asymp \tr \bo \Xi_{k}^{-1} [n\!-\!1]\hat{ \beta}_{llk}\\
 &{\asymp }\tr  \bo T_{li}  \hat{\beta}_{llk},\label{deterministicequivalent2}
\end{align}
where the symbol $\asymp$ relates two infinite sequences and is equivalent to $\xrightarrow[ N \rightarrow \infty]{\mbox{a.s.}}$ denoting almost sure convergence. As far as the derivation is concerned,~\eqref{deterministicequivalent1} follows from~\cite[Lemma B.26]{Bai} as well as in~\eqref{deterministicequivalent2} we have applied the rank-1 perturbation lemma~\cite{Silverstein} and then~\cite[Theorem 1]{Slock}. \endIEEEproof

The deterministic equivalent uplink sum-rate can be obtained by means of the dominated convergence~\cite{Billingsley} and the continuous mapping theorem~\cite{Vaart} as
\begin{align}
R_{k}-\log_{2}\left( 1 +\overline{ \mathrm{SINR}}_{k} \right) \xrightarrow[ N \rightarrow \infty]{\mbox{a.s.}}0.
\end{align}

\section{Numerical Results}
This section provides the representation of the impact of the user mobility on the sum-rate as well as the assessment of the proposed bounds and deterministic equivalent. The scenario under consideration includes $L=7$ cells with 
$K=10$ users per cell. For reasons of comparison, the large-scale fading matrix
$D_{1i}, i = 1, ..., 7$ and the rest parameters, are chosen as in~\cite{Performance_Ngo}\footnote{This large-scale  matrix models practical  channel conditions in hexagonal cells with randomly distributed users under log-normal shadowing $\left( \ln\mathcal{N}\left( 0,8 \right) \right)$ and path loss with exponent equal to $4$.}. Specifically,  $T =196$, $\tau= 10$, $p_{p}=p$, and  $SNR=p$.  Taking as reference cell $1$, the behavior of the spectral efficiency in $bits/s/Hz$ is investigated. In particular, we study the total  spectral efficiency given by
\begin{align}
 R=\left( 1-\frac{\tau}{T} \right)\sum^{K}_{k=1}R_{k},
\end{align}
where $R_{k}$ is the sum-rate of user $k$ in cell $1$ and  $T$ is the coherence time interval (in symbols). Note that $D_{1i}$ is given in terms of $\beta$  expressing the effect of the interference from other cells. It has to be stressed that in all cases the outperformance of the OLR is depicted. In addition, the deterministic approximation of the sum-rate is indistinguishable comparing to the simulation results. It is worthwhile to mention that the sum-rate saturates for high SNR values (interference-limited system), i.e., the system performance cannot be improved by using more power, since increase of the transmit power results also to increase of the interference from other cells.

In Fig.~1, the simulated achievable sum-rate of the OLR along with its deterministic equivalent~\eqref{detequiv} are plotted against the average SNR $p$. In addition, the typical MRC and MMSE decoders are simulated as well. Addressing different interference configurations across different cells, i.e., $\beta=1$ and $\beta=4$, we observe that the OLR presents the best performance, which increases as shown by the zoomed gap, when the intercell
interference increases. The explanation rests on the fact that the correlation between intercell interference and the channel estimate are considered by the structure of the OLR. 

In Fig.~2, the effect of the user mobility by means of the Doppler shift $f_{d}T_{s}$ on various receivers is illustrated by depicting the variation of the sum-rate with the normalized Doppler shift. It can be observed that the Doppler shift extensively limits the performance of the channel. In particular, the sum-rate decreases from almost $9.2~bits/s/Hz$ to zero by following the form of the $\mathrm{J}_{0}\left( \cdot \right)$, while  the Doppler shift increases. Moreover, we observe that the gap between the performances of the various receivers vanishes to zero for values of $\alpha$ greater than $0.25$. Nevertheless, increasing the number of antennas and keeping the other parameters fixed, the sum-rate becomes higher for all values of $f_{D}T_{s}$, as expected.

The tightness of the proposed bounds is demonstrated in Fig.~3, where~\eqref{upperbound},~\eqref{Upper_bound}, and~\eqref{detequiv} are compared with the simulation results. We can easily observe that the proposed approximations are very tight across varying SNR. 
\section{Conclusion}
We have proposed an OLR capable of accounting for not only the detrimental effects of pilot contamination and path-loss, but also the critical time variation of the channel due to the movement of the users. Taking into account the correlation between the channel estimate and the interference, we showed the outperformance of the OLR with comparison to MRC, ZF, and MMSE by deriving closed-form lower and upper bounds of the sum-rate for finite $N$. Moreover, we obtained a deterministic equivalent of the SINR when both the numbers of users and antennas go to infinity with a given ratio under the same channel conditions. Simulations witness the tightness of the bounds as well as the validity of the results. It is worthwhile to mention that the deterministic equivalent provides a tight approximation  of the SINR, allowing us to avoid unbearable time consuming Monte Carlo simulations when the number of BS antennas is very large.

\begin{figure}[t]
    \centering
    \centerline{\includegraphics[width  =0.48\textwidth,height=6.8 cm]{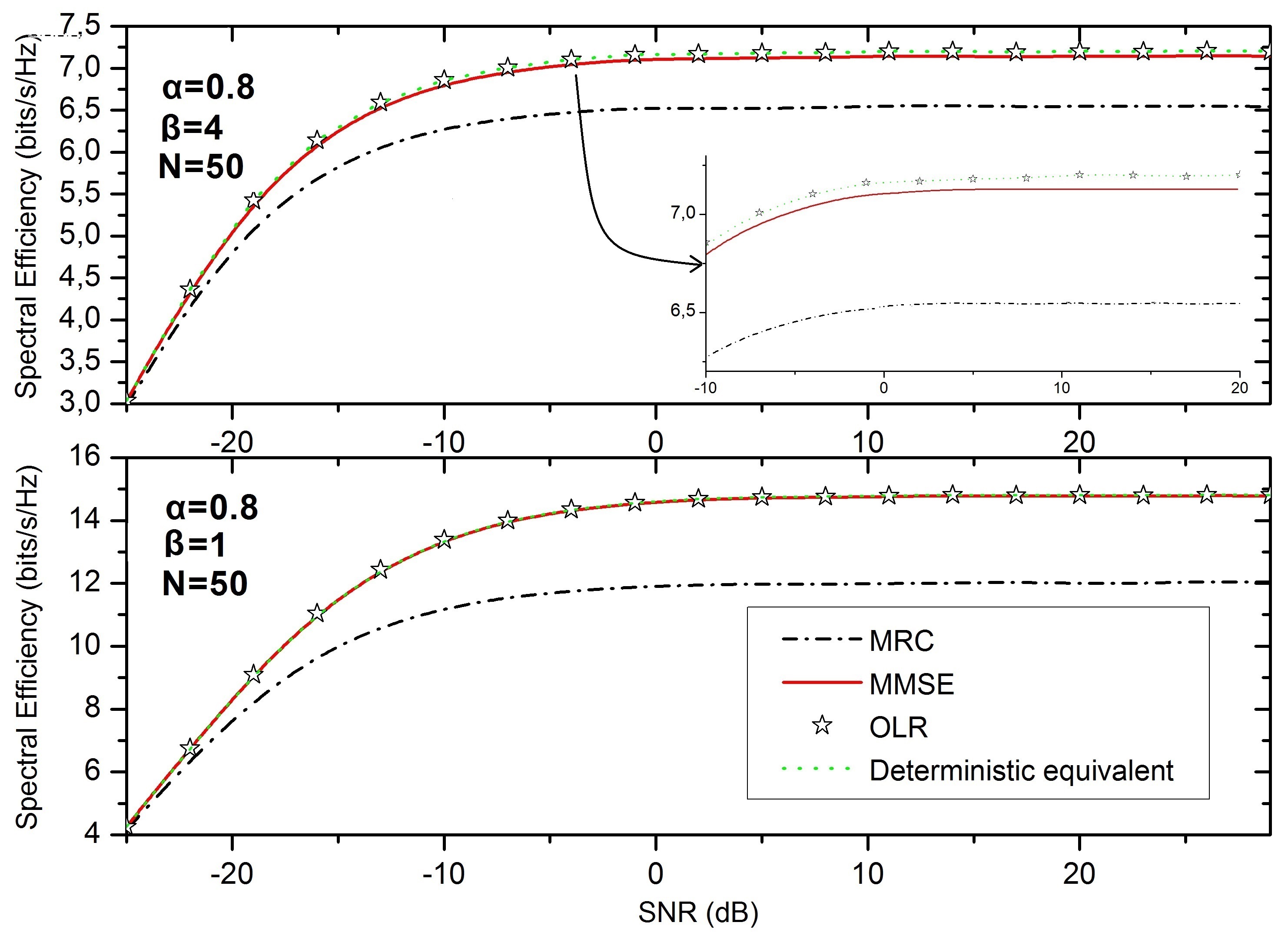}}
    \caption{Sum spectral efficiency  versus $\mathsf{SNR}$ for different interference configurations $(\beta=1$ and $\beta=4)$.}
    \label{fig:1}
\end{figure}

\begin{figure}[t]
    \centering
    \centerline{\includegraphics[width=0.48\textwidth]{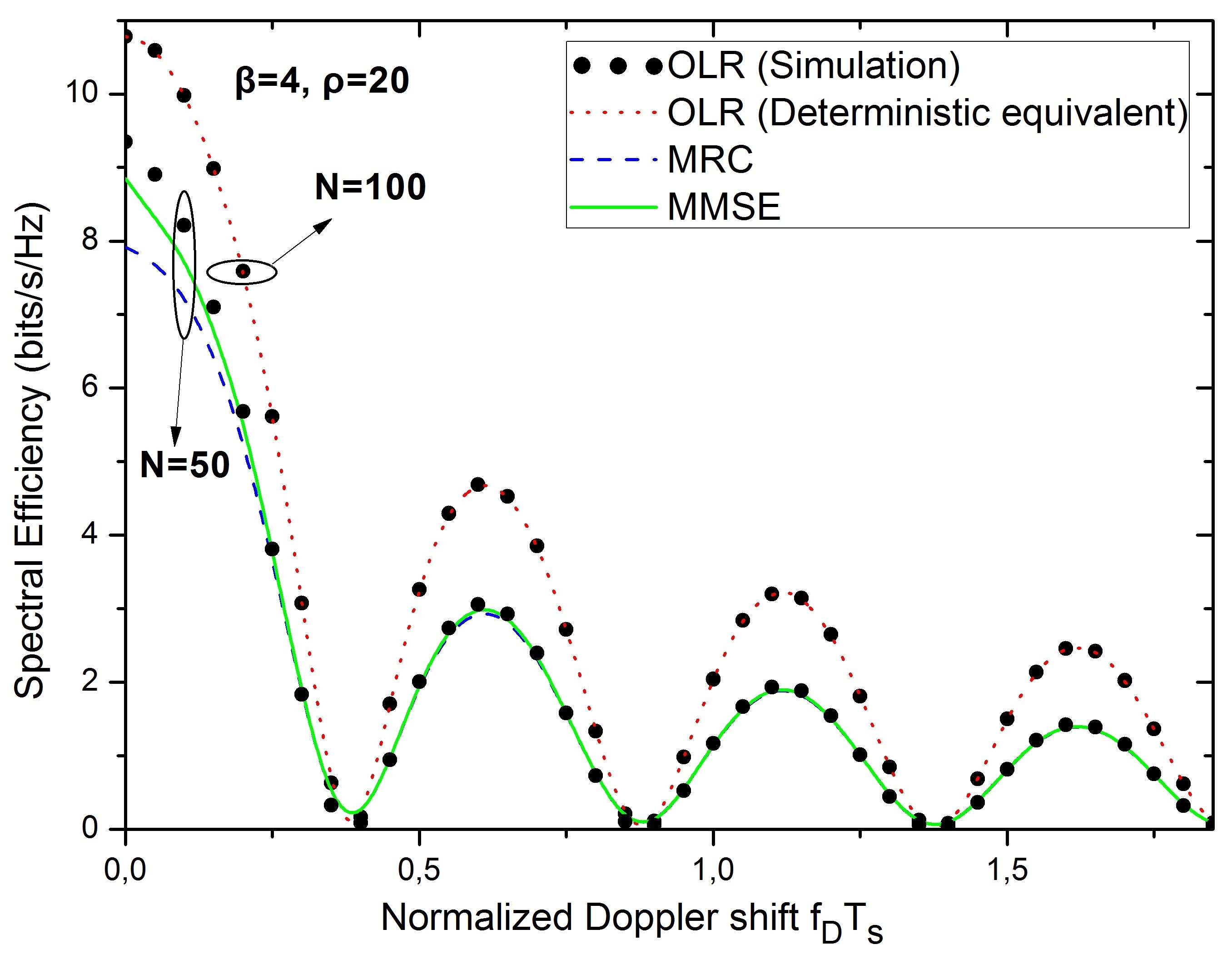}}
    \caption{Sum spectral efficiency  versus normalized Doppler shift for different numbers of BS antennas $(N=50$ and $N=100)$.}
    \label{fig:2}
\end{figure}

\begin{figure}[t]
    \centering
    \centerline{\includegraphics[width=0.48\textwidth]{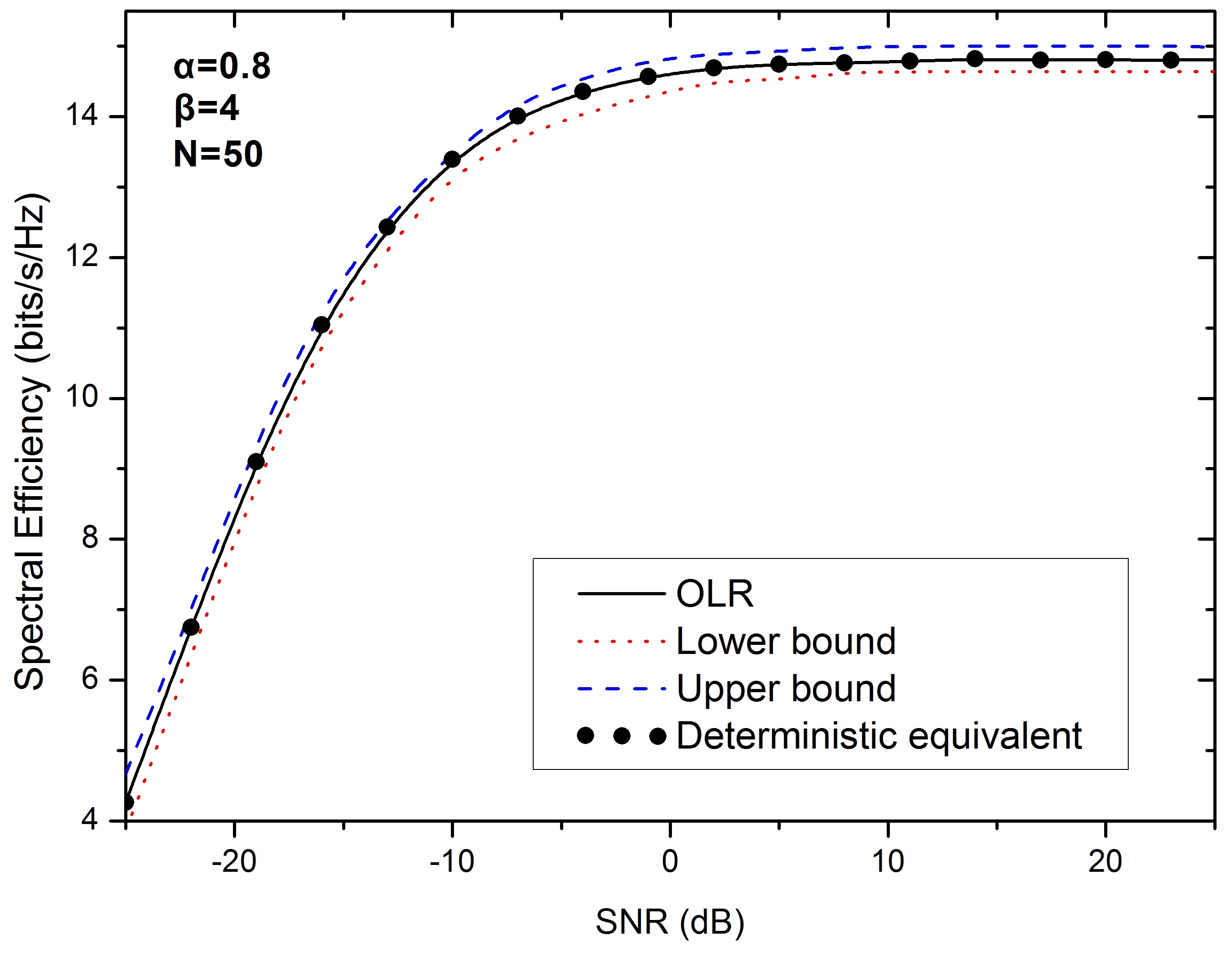}}
    \caption{Comparison of the proposed bounds and the deterministic equivalent with the simulation results.}
    \label{fig:3}
\end{figure}

\end{document}